\documentclass[useAMS,usenatbib]{mn2e}

\include{epsf}
\usepackage{graphicx}
\usepackage{times}
\newcommand{\hi}{H\,{\sc i}}

\newcommand{\msol}{\mbox{${\rm M}_\odot$}}
\newcommand{\hubble}{\mbox{$\rm km\, s^{-1}\, Mpc^{-1}$}}
\newcommand{\kms}{\mbox{$\rm km\, s^{-1}$}}

\newcommand{\mhi}{\mbox{$M_{\rm HI}$}}

\newcommand{\vc}{\mbox{$V_{\rm c}$}}

\newcommand{\mdyn}{\mbox{$M_{\rm dyn}$}}

\newcommand{\vmax}{\mbox{$V_{\rm max}$}}

\title[The velocity function of gas-rich galaxies]{The velocity function of gas-rich galaxies}

\author[M.A. Zwaan et al.]{
\newauthor
M. A. Zwaan,$^{1}$\thanks{email: mzwaan@eso.org} 
M. J. Meyer,$^{2}$
L. Staveley-Smith$^{2}$\\
$^{1}$European Southern Observatory, Karl-Schwarzschild-Str. 2, 85748 Garching
     b. M{\"u}nchen, Germany.\\
$^{2}$International Centre for Radio Astronomy Research, School of Physics, The University of Western Australia, 35 Stirling Hwy, Crawley, WA 6009, Australia
}

\begin{document}

\date{Accepted 2009 December 8,
      Received 2009 August 20}

\pagerange{\pageref{firstpage}--\pageref{lastpage}}
\pubyear{0000}

\maketitle

\label{firstpage}

\begin{abstract}
We measure the distribution function of rotational velocities
$\phi(\vc)$ of late-type galaxies from the HIPASS galaxy catalogue. 
Previous measurements of the late-type velocity function are indirect, derived by converting
the galaxy luminosity function using the relation
between galaxy luminosity and rotation velocity (the Tully-Fisher relation). 
The advantage of HIPASS is that space densities and velocity widths
are both derived from the same survey data. We find good agreement with 
earlier inferred  measurements of $\phi(\vc)$, but we are able to define the
space density of objects with \vc\ as low as 30 \kms. The measured velocity
function is `flat' (power-law slope $\alpha\approx-1.0$) below $\vc\approx 100\,\kms$.
We compare our results with predictions based on $\Lambda$CDM
simulations and find good agreement for rotational velocities in excess of 
100 \kms, but at lower velocities current models over-predict the space density
of objects. At $\vc=30\,\kms$ this discrepancy is approximately a factor 20.
\end{abstract}
 
\begin{keywords}
galaxies: luminosity function, mass function --
dark matter --
radio lines: galaxies
 
\end{keywords}

\section{Introduction}

The galaxy luminosity function is one of the best-known
tools for comparing theoretical models of galaxy formation with
the observable universe. Modern large scale cosmological simulations, based 
on the hierarchical clustering of gravitationally interacting particles, 
need to reproduce the space density of galaxies seen in the observed luminosity function.
However, in reality such a comparison is complicated because
galaxy luminosities are not a direct output of these simulations. 
Galaxy luminosity is non-trivially linked to halo mass through poorly understood
processes such as star formation efficiency, feedback, and gas accretion. 
Recent semi-analytical models  attempt to tie the $N$-body simulations to 
observable quantities by applying physically motivated or empirical recipes for star formation, gas accretion, AGN feedback, etc. \citep[e.g.,][]{Baugh2005a,Bower2006a,DeLucia2007a,Obreschkow2009c}. These models
produce a range of measurable galaxy properties, such as luminosity
in different bands, gas masses, and star formation rates, and the distribution
functions of these quantities can be tested against observations.
Although such semi-analyical models are increasingly more successful in reproducing the
observable universe \citep[see review by][]{Baugh2006a}, the method inherently implies 
a very indirect way of comparing galaxy space densities. A much more
direct test of the simulations can be made if the properties of simulated haloes 
are compared directly with observable quantities. 
Such a comparison is possible through the galaxy circular velocity
function $\phi(\vc)$, which is defined as the number density of galaxies
as a function of their rotational velocity.

The galaxy circular velocity traces the total dynamical mass of a galaxy,
which is dominated by its dark matter content. Therefore, the measured
velocity function should match that of dark haloes in simulations. However, 
it should be realized that baryons still potentially affect the comparison 
in two unrelated ways. First, for surveys to detect the galaxies
and include them in samples that are used to create the velocity function,
the galaxies need to have a minimal detectable amount of baryons in the component and  phase being searched for. Whether 
in an individual galaxy these baryons are primarily in cold gas, in old stars or young stars, may
affect this galaxy's detectability. Second,  baryon physics can significantly
affect the density structure of dark matter haloes and consequently have an
effect on its rotational velocity. Especially in the inner regions of galaxies the
baryonic component can contribute significantly to the rotational velocity and
in some cases the baryons can dominate the dynamics in the central regions.
 Therefore, although the velocity function provides a
much more direct tool to compare simulations and observations than
galaxy luminosity functions, these two limitations should be kept in mind
in making the comparison. 

There have been a number of attempts in the past to construct the galaxy velocity function by using published luminosity functions and
then converting these into velocity number density distributions using
scaling relations such as the Tully--Fisher relation (for late-type galaxies) and the Faber--Jackson relation or the fundamental plane (for early-type galaxies). \citet{Shimasaku1993a} and \citet{Gonzalez2000a} apply this method on an array of optical surveys and find consistent  velocity functions (within a factor of 2), with a Schechter-like distribution, but with a steeper than exponential decline at the high velocities. As is stressed by \citet{Kochanek2001a} and demonstrated by \citet{Sheth2003a}, this simple conversion from a luminosity function via a scaling relation can lead to erroneous measurements of $\phi(\vc)$ if the scatter on the scaling relations is not taken into account. \citet{Kochanek2001a} mention a number of other systematic problems with this method of measuring $\phi(\vc)$, prime among which are the errors in morphological classification of galaxies and calibration of the magnitude scale. They put special effort in reducing these systematic uncertainties and measure a non-parametric $\phi(\vc)$. \citet{Sheth2003a} measure $\phi(\vc)$ for early type galaxies from SDSS taking full account of the scatter in the relation between luminosity and velocity, \citet{Chae2008a} takes a similar approach for SDSS and 2dF galaxies, and \citet{Choi2007a} use volume-limited SDSS subsamples to construct a velocity function for early types.

In this paper we present the first measurement of the velocity function of late-type galaxies that is not dependent on scaling relations. We base this measurement on the \hi\ Parkes All Sky Survey (HIPASS), a blind 21-cm emission line survey covering the whole southern sky up to declination $+25^\circ$. From the region south of declination $+2^\circ$ a catalogue of 4315 \hi-selected galaxies has been extracted as presented in \citet{Meyer2004a} and \citet{Zwaan2004a}. The advantage of this large galaxy sample is that the velocity width $W$ is one of the data products of the original survey data. Therefore, after doing the appropriate volume corrections and correcting the velocity widths for inclination, the velocity function can be measured directly without taking the step of measuring \vc\ separately, or assuming a global scaling relation for all galaxies. The HIPASS catalogue is 99 per cent complete at a peak ßux of 84 mJy and an integrated ßux of 9.4 $\rm Jy km s^{-1}$. This integrated flux limited translates to a minimal detectable \hi\ mass of $5.5\times 10^7 \msol$ at a distance of 5 Mpc.

The paper is organized as follows. In section 2 we present the galaxy sample we use in our analysis, summarize the method for calculating space densities and  discuss the observed velocity width function.  Velocity functions for late-type galaxies are presented and compared with previous inferred velocity functions in section 3. We also discuss the contribution from early-type galaxies. In section 4 we compare our results with those of recent $\Lambda$CDM simulations. Section 5 discusses some observational biases that could affect the evaluation of the velocity function. Finally, we summarize our results in section 6.
For distance dependent quantities we use $H_0=70~\hubble$ and define $h_{70}=H_0/(70~\hubble)=1$.


\section{The HIPASS velocity width function}

\subsection{Calculating space densities}
In \citet{Zwaan2003a} we derived a bivariate stepwise maximum likelihood (2DSWML) technique, which solves for the space density of objects as a function of \hi\ mass and velocity width simultaneously. We specifically designed this method for a peak-flux-limited sample, but in \citet{Zwaan2004a} we demonstrated how the method can be adjusted to produce reliable space densities for samples constructed using different selection criteria. In that paper we also showed how to use the 2DSWML technique to calculate for each galaxy individually an effective volume $V_{\rm eff}$.  These values are now the maximum likelihood equivalents of \vmax\ for the standard $\Sigma 1/\vmax$ method \citep{Schmidt1968a} and summing the values of $1/V_{\rm eff}$ in bins of \hi\ mass gives the original 2DSWML solution for the \hi\ mass function. The normalization of space densities is retained by equating the mean galaxy density $\bar{n}$ to the integral over the \hi\ mass function. After this normalization is applied to $V_{\rm eff}$, these values represent true volumes. Any distribution function of a parameter measured for the HIPASS sample can now be determined by a binned summation of $1/V_{\rm eff}$. We refer to \citet{Zwaan2004a} for a more detailed description of the method.

\subsection{The velocity width function}
We use the velocity widths $W$ measured at 50 per cent of the peak flux density of the \hi\ profile.  In principle, for high signal to noise data, the widths measured at 20 per cent of the peak flux density will provide more direct measurements of the full rotation, but we prefer the 50 per cent values
because these latter values are less sensitive to noise in the 21-cm spectra.
\citet{Meyer2008a} have shown that the scatter in the Tully-Fisher relation is smaller when using 50 per cent measurements instead of those measured at 20 per cent. The observed line widths 
are corrected to their rest-frame values (relativistic effect), instrumental broadening and turbulent motion, all according to standard procedures. A full description of the corrections applied to the observed velocity widths is described in \citet{Meyer2008a}. Henceforth, we refer to the velocity width corrected for instrumental and relativistic broadening as $W$. Note that for the calculation of rotational velocities an additional correction for turbulent broadening is applied.

\begin{figure}
\begin{center}
\includegraphics[width=7.0cm,trim=1.2cm 4.3cm 1.5cm 0cm]{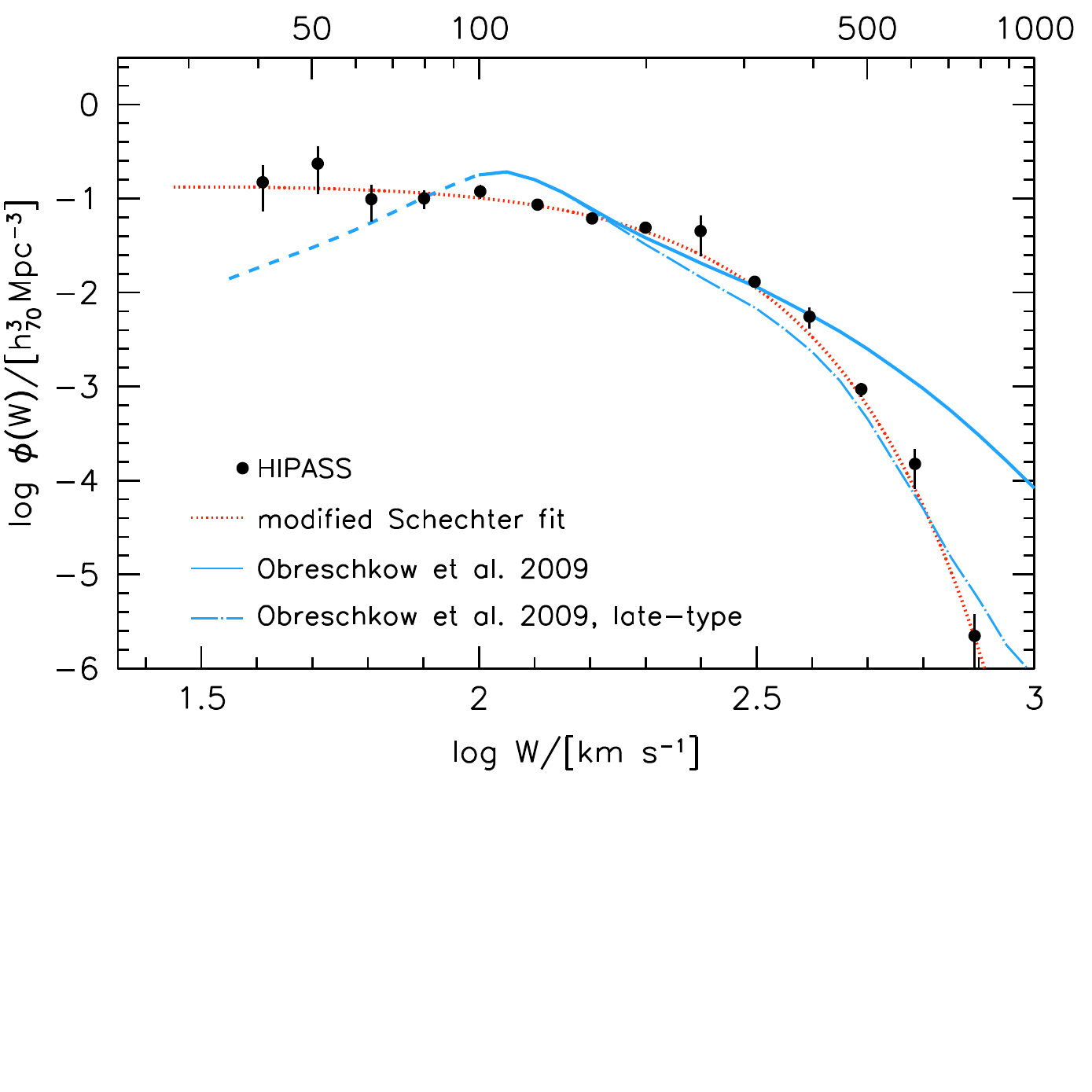}
\caption{\label{velf_W.fig} The space density of local gas-rich galaxies as a function of their observed profile widths corrected for relativistic broadening and instrumental broadening. The short-dashed line is a modified Schechter fit to the data points (see text). The solid line is the distribution function based on the  \citet{Obreschkow2009c} simulations, while the long-dashed line represents this sample limited to late-type galaxies only. }
\end{center}
\end{figure}

Fig.\ref{velf_W.fig} plots the distribution function of \hi\ profiles widths, $\phi(W)$. This function can be readily calculated by summing $1/V_{\rm eff}$ in bins of $W$.  Error bars represent uncertainties in the measurements determined by quadratically summing $1/V_{\rm eff}$ within each bin. This estimate gives an indication of the pure statistical uncertainty, which mostly ignores effects due to local large scale structure. All 4315 HIPASS detections presented in \citet{Meyer2004a} are represented in this diagram as no optical identifications are required for constructing this function. The error on $W$ measurements is $\sigma_W=7.5\, \kms$, independent of profile width \citep{Zwaan2004a}.

To provide a parametric description of this distribution, we fit it with a modified Schechter 
function of the form
\begin{equation}
\phi(W) dW = \phi_* (\frac{W}{W_*})^\alpha \exp[-(\frac{W}{W_*})^\beta]  \frac{\beta}{\Gamma(\alpha/\beta)} \frac{dW}{W},
\end{equation}
where $\Gamma$ is the gamma-function.
In the next section we give a physical motivation for this parameterization of a velocity function. We find
a good fit to the measured data points with the following values: 
$\phi_*=0.67\, h_{70}^3 {\rm Mpc}^{-3}$,
$\log W_*/\kms=2.21$,
$\alpha=0.10$,
and $\beta=1.56$.

Although no true physical parameter is represented in this diagram, it does present a straightforward comparison to simulated galaxy samples. Since it does not depend on cross-matching with optical catalogues, it shows the distribution function of observed profile widths of a purely \hi-selected sample. This function is relevant for calculating the detection efficiency of future \hi\ 21-cm surveys, as the ability to separate 21-cm signals from the noise is a strong function of profile width. 

An interesting comparison is with the work of \citet{Obreschkow2009c}, who calculate cold gas properties for the semi-analytic galaxy catalogues of \citet{DeLucia2007a} based on the $N$-body simulations of cold dark matter by \citet{Springel2005a}. Velocity widths are assigned to simulated galaxies by making assumptions on the rotation curves and gas distributions. We refer to \citet{Obreschkow2009c} for details on how velocity widths are calculated. 
The solid line in Fig.~\ref{velf_W.fig} shows the distribution function of $W$ from these simulations. Below $W=100~\kms$ we draw a dashed line to indicate that the simulations are incomplete in that range due to the limited mass resolution. For the $W$ measurements from  \citet{Obreschkow2009c} we used the raw $W_{50}$ values multiplied by $\sin i$.
In the range $100~\kms<W<400~\kms$ the HIPASS points match the \citet{Obreschkow2009c} points fairly well, although the simulations over-predict the space density of galaxies just above the mass resolution limit.
Toward higher velocities the observations show an exponential decline, whereas the observations display a power-law distributions. Clearly, the simulations predict too many very broad profile ($W>400~\kms$) galaxies compared to what HIPASS detected. The \hi\ mass function of the  \citet{Obreschkow2009c} catalogue also shows an overproduction of very massive galaxies (their Fig. 3), but the difference in our 
Fig.\ref{velf_W.fig} is much larger. To investigate this discrepancy further, we also plot in the same figure the simulated distribution of late-type galaxies only, and find that the match the observed distribution (without any morphological type cut applied) is much better. Therefore, it appears that the \citet{Obreschkow2009c} simulations overproduce a population of gas-bearing massive early-type galaxies that is not observed in HIPASS. 

\section{A type-dependent rotational velocity function}
\subsection{A velocity function for late-type galaxies}

The next step is to determine the distribution of true rotational velocities. These \vc\ values we calculated by applying an inclination correction to the projected velocity spread of the \hi\ gas: $\vc = W (2\sin\ i)^{-1}$. Henceforth, we refer to the rotational velocity \vc\ as the velocity width corrected for turbulence, instrumental broadening, inclination and corrected to rest-frame profile widths,
as detailed in \cite{Meyer2008a}. We note that  \vc\ should be regarded as the velocity corresponding to the maximum of the rotation curve as this is essentially the velocity that is probed by integrated \hi\ velocity profiles. In principle, more accurate measurements of galaxy halo masses could be achieved my making use of $V_{\rm flat}$, the velocity corresponding to the flat part of the rotation curve, but high resolution 21cm synthesis observations are required to make such measurements. \citet{Verheijen2001b} discusses the use of \vc\ and $V_{\rm flat}$ in the Tully-Fisher relation and their relation to halo mass. 

The required inclinations cannot be measured from the HIPASS data directly because the Parkes Telescope beam at 21-cm of $13'$ (FWHM) does not spatially resolve most galaxies.
\citet{Doyle2005a} searched for optical counterparts to the HIPASS detections
and found unique optical identifications for 2646 (62 per cent) galaxies. For these
galaxies \citet{Doyle2005a} list $b_J$-band axis-ratios measured from the SuperCOSMOS
data. Inclinations are calculated from the axis ratios $q$ and corrected for
internal thickness of the disks $q_0$ using the relation
\begin{equation}
\cos^2 i = \frac{q^2-q_0^2}{1-q_0^2},
\end{equation}
where we adopt $q_0=0.2$. It has been argued that $q_0$ should depend on galaxy morphology \citep{Giovanelli1997a} but we follow the recommendation of \citet{Tully2009a} that a consistent use of $q_0$ is important for the avoidance of systematic errors. We note that \citet{Meyer2008a} found that inclinations derived from the $b_J$ measurements resulted in lower Tully-Fisher scatter that those from the 2MASS $J$-band measurement.

Selecting galaxies on the basis of their \hi\ content naturally results in a sample weighted toward gas-rich galaxies. However, this does not imply that early-type galaxies (S0/E) are missing from the catalogue. In fact, some recent studies have shown that 70 per cent of early-type galaxies have detectable amounts of \hi\ with masses exceeding a few times $10^6\,\msol$ \citep[e.g.,][]{Morganti2006a,Oosterloo2007a}. The \hi\ morphologies range from 
small clouds to large, regular low column density disks. Furthermore, in \citet{Zwaan2003a} we have measured the \hi\ mass function of early-type galaxies, based on an initial sample of 1000 bright HIPASS galaxies, indicating that early-types are present in 21-cm survey catalogues.
However, in the analysis of the HIPASS data we wish to exclude the early-type galaxies for two reasons. First, even though a large fraction of early-types has detectable \hi, we are certainly  missing a significant fraction of   those early-types that are gas free or sufficiently gas poor to be absent from the HIPASS sample. Secondly, \hi\ measurements do not provide reliable velocity dispersions for early-type galaxies. As already stated,  early-type \hi\ morphologies are not predominantly ordered disks that trace the full potential of the galaxies. This fact is also supported by the observation that
 \hi\ masses do not correlate with early-type luminosity or mass \citep{Serra2008a}.
Optical surveys such as the SDSS, completed with optical spectroscopy are much better equipped to measure the distribution function of early-type velocity dispersions \citep{Sheth2003a,Choi2007a, Chae2008a}. To cull the early-type galaxies from the sample we rely on the morphological classifications as listed by \citet{Doyle2005a}. This takes out 11 per cent of the galaxy sample. 

We also exclude all galaxies with measured inclinations $i<45^\circ$ to avoid introducing large uncertain inclination corrections that may affect the shape of the velocity function. This amounts to 30 per cent of the sample. To account for the excluded galaxies, the final velocity function is corrected with a factor $1/(1-f)$ where $f$ is the fraction of excluded galaxies.

\begin{figure}
\begin{center}
\includegraphics[width=7.0cm,trim=1.2cm 0.7cm 1.5cm 0cm]{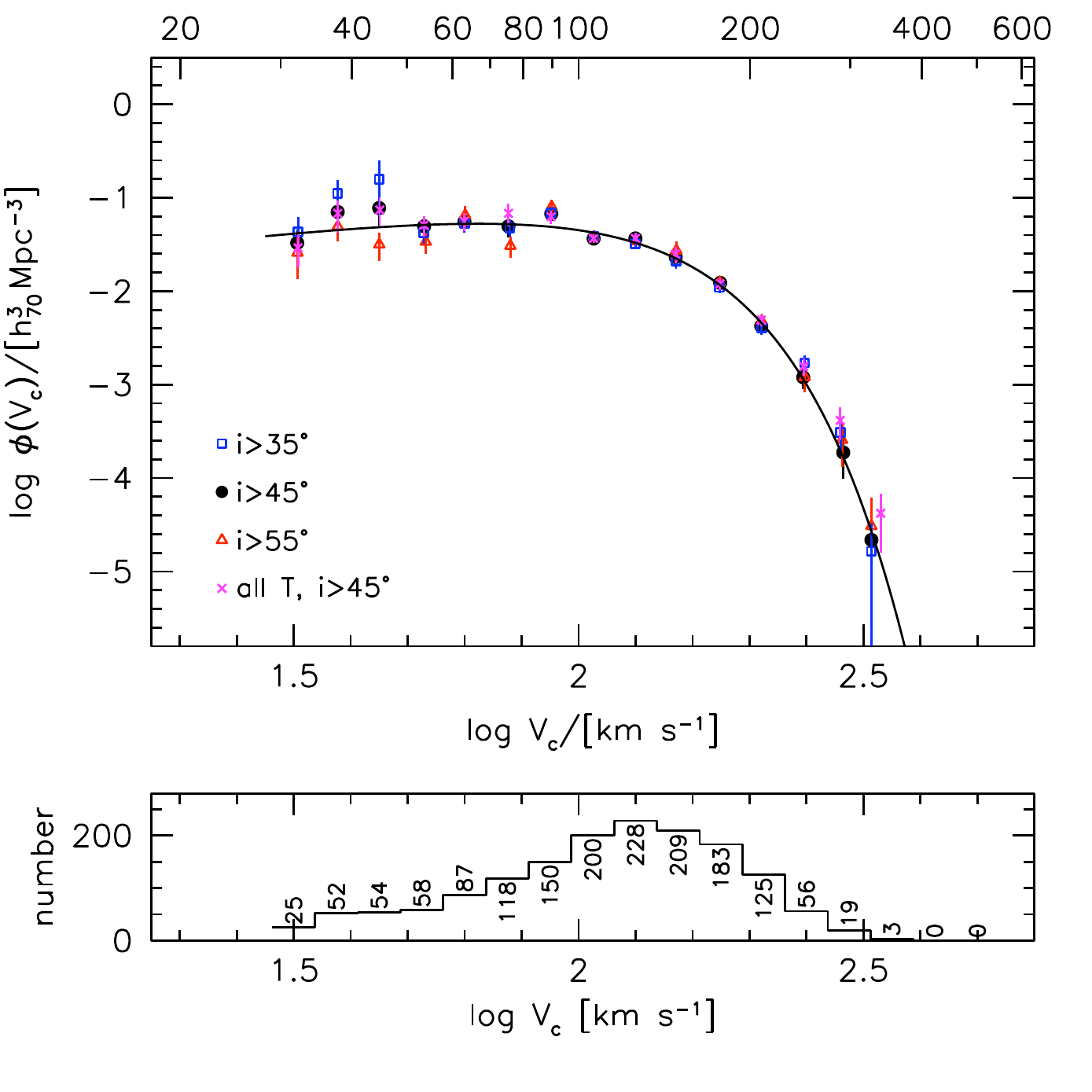}
\caption{\label{velf_bias.fig} {\em Top panel\/:} The HIPASS late-type
velocity function $\phi(\vc)$. The solid circles show $\phi(\vc)$ for
morphological types Sa and later  and inclinations
$i>45^\circ$. The other symbols show the effects of choosing different
inclination cut-offs and including early-type galaxies.  The solid line represents
the best-fit modified Schechter function (see text).
{\em Bottom panel:\/}
Histogram of rotational velocities for our sample of HIPASS galaxies
with $i>45^\circ$ and types Sa and later.  }
\end{center}
\end{figure}

In Fig~\ref{velf_bias.fig} the measured velocity function is shown as solid circles, whereas the open symbols indicate the effect of changing the inclination cut-off from 45 to 35 and 55 degrees. The effect of changing the inclination cut-off is not noticeable for \vc\ above approximately 100 \kms, but at lower \vc\ the changing cut introduces some scatter. There is a weak trend of space densities dropping with a higher inclination cut, and increasing with a lower inclination cut, although the error bars are too large to make a conclusive statement: the mean space density of objects with $\vc<80\,\kms$ changes less than $1\sigma$ when the inclination cut is varied from 45 to 35 or 55 degrees.

Also in  Fig~\ref{velf_bias.fig}, we show the effect of ignoring the morphological type selection by plotting the velocity function for all morphological types.  The effect is only detectable beyond $\vc>200\,\kms$, but even there early-type galaxies make a very minor contribution to the space densities.

Similar to the distribution function of $W$, we fit the rotational velocity function with a modified Schechter function of the form
\begin{equation}
\phi(\vc) d\vc = \phi_* (\frac{\vc}{\vc_*})^\alpha \exp[-(\frac{\vc}{\vc_*})^\beta]  \frac{\beta}{\Gamma(\alpha/\beta)} \frac{d\vc}{\vc},
\end{equation}
where $\phi_*$ is the space density of galaxies near the `knee' of the distribution, $\vc_*$ is the characteristic rotational velocity, $\alpha$ is the low-velocity power-law index, and $\beta$ is the high-velocity power-law index. The latter parameter $\beta$ enters the equation when the traditional Schechter luminosity function is transformed to a velocity function using a relation between luminosity $L$ and rotational velocity, e.g., the Tully-Fisher relation: $L\propto \vc^\beta$. In reality, the value of $\beta$ in the parametrized velocity function is not equal to the Tully-Fisher power law because the scatter in the Tully-Fisher relation affects the way in which the luminosity function is transformed to $\phi(\vc)$, and because the Tully-Fisher power law changes as a function of luminosity and morphological type. These effects have been described in detail by e.g., \citet{Sheth2003a}. A traditional Schechter function would provide a very poor fit to the measured velocity function because the high end exponential drop off does not decline fast enough to fit the data points.
We find a satisfactory fit to the data with the following parameters:
$\phi_*=(0.061\pm 0.027)\, h_{70}^3 \,{\rm Mpc}^{-3}$,
$\log \vc_*/ \kms = 2.06\pm 0.09$,
$\alpha=0.66\pm 0.47$, and
$\beta=2.10\pm0.37$. It should be noted that errors on individual parameters are very large because of the strong covariance between all fitting parameters. In particular, the dimensionless covariance between $\log \vc$ and $\beta$ is $r(\log \vc,\beta)=0.99$ and that between $\phi_*$ and $\alpha$
is $r(\phi,\alpha)=-0.97$. The lowest covariance value is $r(\phi_*,\beta)=0.78$. By just fitting the low velocity slope ($\vc<100\,\kms$) of the distribution, we find $\alpha=-0.90\pm 0.25$.

\begin{figure}
\begin{center}
\includegraphics[width=7.0cm,trim=1.2cm 4.3cm 1.5cm 0cm]{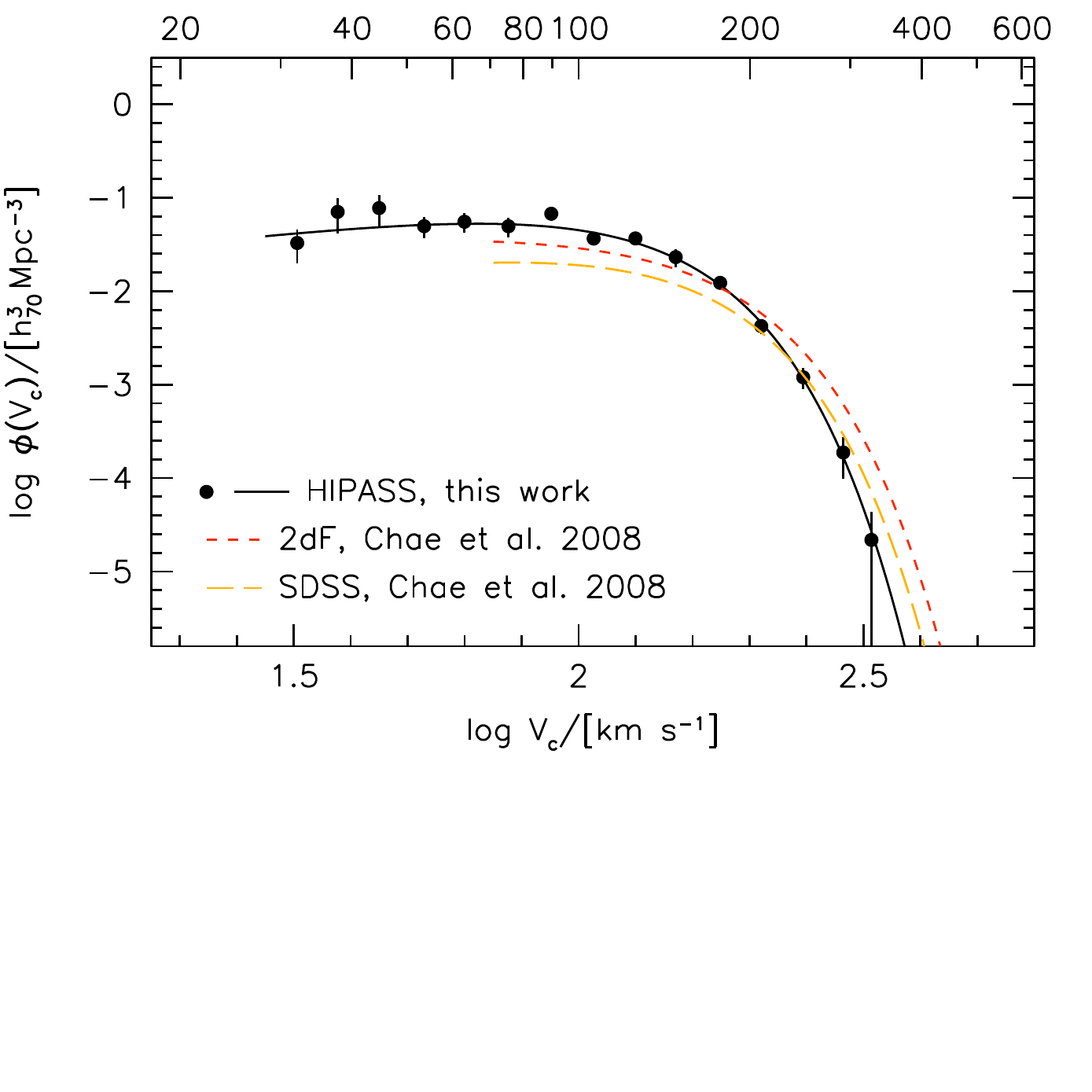}
\caption{\label{velf_late.fig} The HIPASS late-type velocity function $\phi(\vc)$ as indicated by the circles with error bars. The solid curve is a modified Schechter function fit. The two dashed curves show the late-type velocity function based on the 2dF data (upper curve) and SDSS data (lower curve) as calculated by \citet{Chae2008a}. }
\end{center}
\end{figure}

Fig.~\ref{velf_late.fig} reproduces the HIPASS velocity function and compares our result with recent estimates of the late-type velocity function based on the 2dF and SDSS galaxy redshift surveys from \citet{Chae2008a}.  In that study, published luminosity functions were adopted for the SDSS late-type galaxies from \citet{Choi2007a} and for 2dF from \citet{Croton2005a}. Next, they Monte Carlo simulate a large sample of galaxies, based on these luminosity functions and 
the Tully-Fisher relation, taking into account the intrinsic scatter. For the Tully-Fisher relation, the measurement by \citet{Pizagno2007a} was adopted, which is based on $H\alpha$ rotation curves of a sample of 162 SDSS galaxies. The velocity functions are found by summing the simulated galaxies in bins of \vc. In Fig.~\ref{velf_late.fig}, the upper dashed curve shows the result based on the 2dF luminosity function and the lower dashed curve is based on SDSS. Discrepancies between the two function are indicative of the systematic errors introduced by converting galaxy luminosities between different wave bands that are used in the luminosity function and Tully-Fisher relation measurements, different definition of morphological type (based on spectroscopy or colour), and differences in the 2dF and SDSS galaxy luminosity functions \citep[as discussed in e.g.,][]{Liske2003a}.

Given the very different galaxy samples and methods for calculating the velocity function, we
find reasonable agreement between the {\em directly measured} HIPASS velocity function and the {\em inferred} 2dF and SDSS velocity functions. At lower velocities, we find slightly higher space densities compared to the indirect measurements and at higher velocities we find slightly lower values.
The most significant difference between our measurement and previous estimates is that we can determine the space densities of galaxies down to much lower values of \vc: the HIPASS velocity function goes down to $\vc \approx 30\,\kms$, whereas the previous measurements only reach $\vc\approx 70\,\kms$. The measured velocity function is remarkably `flat' (power-law slope $\approx -1$) down to the lowest rotational velocities.


\subsection{The contribution from early-type galaxies}
Because 21-cm emission is not a good tracer of the gravitational potential of early-type galaxies, we have so far excluded these galaxies from our analysis. It would, however, be interesting to derive the total local universe velocity function of all galaxy types.  To put early and late-type galaxies on the same \vc\ scale, we adopt a conversion from velocity dispersion $\sigma$ to rotational velocity: $\vc=\sqrt2\sigma$. This conversion is based on the assumption that the mass distribution of an elliptical galaxy can de modeled as a singular isothermal sphere. 
\citet{Desai2004a} test this assumption using literature measurements of velocity dispersion and 
optical circular velocities in the flat regions of the rotation curves
and find $\vc\approx 1.54\sigma$, illustrating the uncertainty in estimating $\vc$ values for early-types.

\begin{figure}
\begin{center}
\includegraphics[width=7.0cm,trim=1.2cm 4.3cm 1.5cm 0cm]{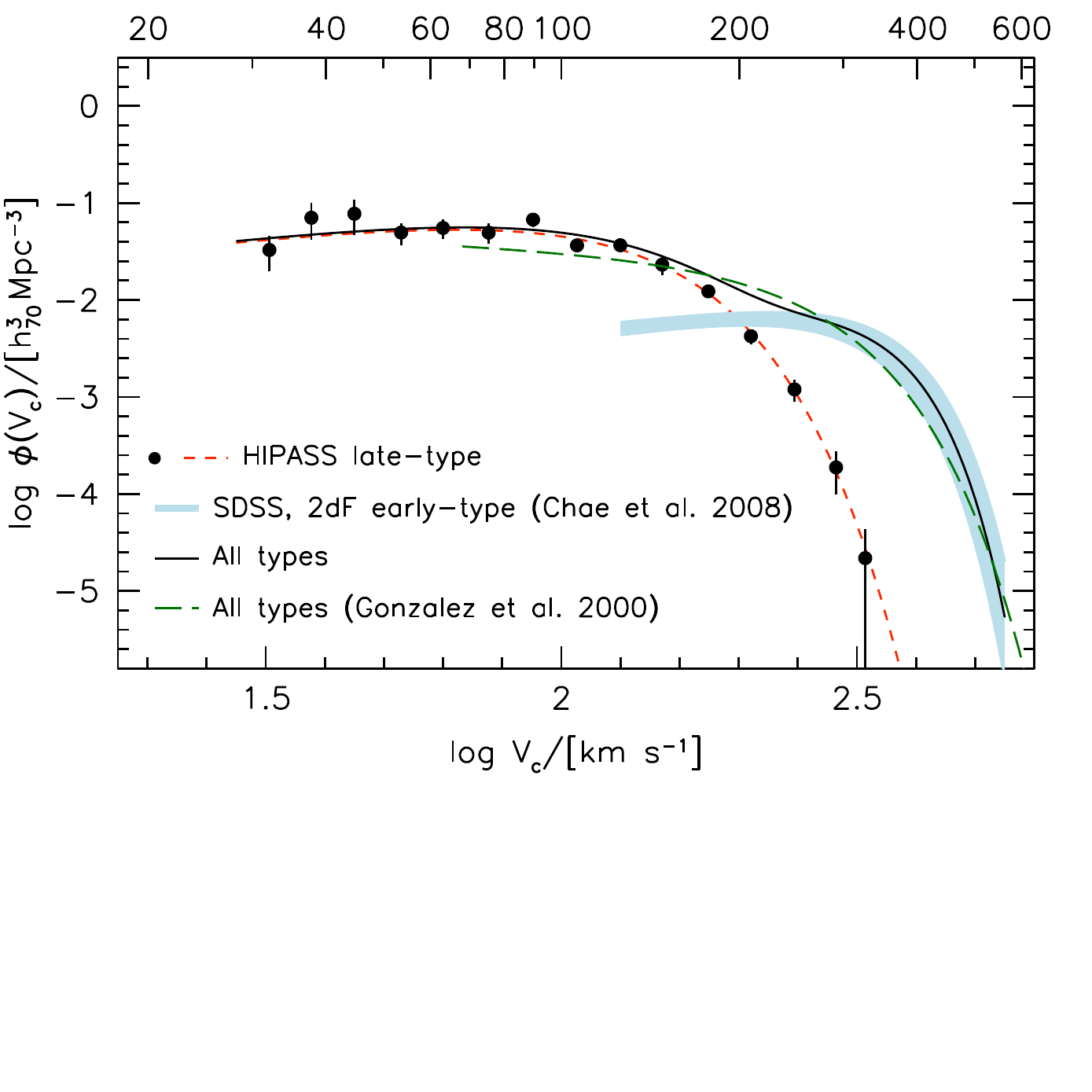}
\caption{\label{velf_all.fig}  The points with error bars show the HIPASS late-type velocity function with a modified Schechter function fit indicated by a short-dashed line. The shaded curve represents the early-type velocity function derived from 2dF data (upper boundary) and SDSS data (lower boundary) as estimated by \citet{Chae2008a}. The solid curve is the summed velocity function for all morphological types. The \citet{Gonzalez2000a} all-types result is represented by a long-dashed line.
 Late-type galaxies dominate the statistics below $\vc\approx 200 \,\kms$.}
\end{center}
\end{figure}

The most recent estimates of early-type velocity functions are those by \citet{Chae2008a}, based on 2dF and SDSS using a method similar to that described for the late-type galaxies. Fig.~\ref{velf_all.fig} shows these measurements as indicated by a shaded thick curve, of which the upper boundary corresponds to the calculation based on 2dF data and the lower boundary corresponds to the SDSS. Other recent calculations of the SDSS velocity function have been presented by \citet{Sheth2003a} and \citet{Choi2007a}, the latter being based on direct measurements of velocity dispersions. However, this study may suffer from incompleteness at low velocity dispersions. As shown by \citet{Chae2008a}, the different space density measurements  of early-type galaxies around the peak of the distribution ($\vc\approx 300\,\kms$) agree very well, but below that point the slopes of $\phi(\vc)$ vary between the different studies. Issues with completeness, both related to incomplete dispersion measurements and missing low luminosity early-types, are probably causing the discrepancies. See also \citet{Mitchell2005a} for a discussion on inferred and measured early-type velocity functions. 

The solid curve in Fig.~\ref{velf_all.fig} shows the summed velocity function for all morphological types, adding the HIPASS late-type curve and the 2dF/SDSS-based early type curve of \citet{Chae2008a}.
For comparison, we also show the velocity function from \citet{Gonzalez2000a} based on SSRS2 data. Their result is in good agreement with later calculations by \citet{Kochanek2001a}, although these authors do not give a parametrized description of their velocity function. The overall consistency between the HIPASS plus 2dF/SDSS velocity function with the earlier work of \citet{Gonzalez2000a}  is reassuring. Again, HIPASS adds important data points at the low \vc\ end of the velocity function.

Fig.~\ref{velf_all.fig} illustrates that the galaxy velocity function is dominated by late-type galaxies below $\vc\approx 200 \,\kms$, whereas early-type galaxies take over above that point. It should be noted that the early-type measurements are based on samples having relatively high luminosity cut-offs. Dwarf elliptical (dE) galaxies are therefore largely missing from the analysis. In field regions late type-spirals and irregular (Irr) galaxies dominate the space densities at the low mass end. This is both apparent from optical surveys \citep{Binggeli1988a} and \hi\ surveys \citep{Zwaan2003a}. Dwarf ellipticals only become significant contributors to the space density in cluster environments. The evidence from \hi\ mass function studies is that 21cm surveys fully recover the the faint-end slope of the luminosity function \citep{Zwaan2001d}. Therefore the contribution from gas-less early-type dwarfs is marginal in the volume sampled by HIPASS. 
\citet{Desai2004a} construct velocity functions for cluster galaxies, and find that their function is steeper than that in the field, which is also attributed to a contribution of low mass early-type galaxies, which are relatively scarce in a large volume of space.

\section{Comparison with theory}

The  cold dark matter model of galaxy formation with a cosmological constant ($\Lambda$CDM) makes robust predictions for the mass function of dark matter haloes. Cosmological simulations determine the mass function slope with high accuracy, giving $\psi(M) \propto M^{-1.9}$ \citep[e.g.,][]{Sheth2001a,Sheth2002a,Giocoli2008a,Tinker2008a,Boylan-Kolchin2009a}. To convert this function into the distribution function of individual galaxies, halo occupation distribution (HOD) has to be applied \citep{Peacock2000a,Cooray2002a}. The number density of galaxies may be higher than that of dark matter haloes since each dark matter halo can contain a number of galaxies. 

$\Lambda$CDM models generally predict that the total dynamical galaxy  mass \mdyn\ relates to rotational velocity \vc\ as $\mdyn \propto \vc^\beta$. In our definition of rotational velocity,  \vc\ is comparable to the maximum rotational velocity, for which definition \citet{Bullock2001a} found 
that  $\beta\approx 3.4$. Applying this conversion to the power-law mass function slope of $\alpha_{\rm M}=-1.9$, we can estimate the power-law slope $\alpha$ of the velocity function
\begin{eqnarray}
\phi({\vc}) d\vc &=& \psi({M}) dM\\
&=&\vc^{\alpha_M\beta} \left ( \frac{dM}{d\vc}\right ) d\vc\\
&=&\vc^{\alpha_M\beta+\beta-1}d\vc =\vc^{\alpha}\,d\vc, \label{slope.eq}
\end{eqnarray}
finding that power-law slope of $\phi(\vc)$ should be approximately $\alpha=-4.0$. Of course, this simple estimate ignores many subtleties of HOD and baryon physics, but is does illustrate that theory probably predicts a low-\vc\ slope much steeper than $\alpha\approx-1$, which we find for the HIPASS velocity function. 

The clearest manifestation of this problem with $\Lambda$CDM is the so-called substructure problem: around the Milky Way Galaxy
$\Lambda$CDM predict hundreds of low-mass satellites, although only an order of magnitude fewer objects are detected \citep{Klypin1999a,Moore1999a}. Recent surveys identify a growing number of satellites around the Galaxy and M31 \citep[see][for some examples]{Zucker2006a,Ibata2007a}, but still not in sufficient quantities to  alleviate the problem. By accounting for baryonic physics, \citet{Simon2007a} manage to reduce the missing fraction of satellites to approximately a factor of 3, but baryon physics are extremely complicated and uncertain in these dense environments to the extent that reconciliation between theory and observations is presently too model dependent. We note however that \citet{Koposov2009a} demonstrate that models with a constant stellar mass fraction and strong suppression of star formation before reionization in haloes with $\vc<10\kms$ can produce the observed satellite counts.

The range of masses that we are probing with HIPASS is clearly higher than that of the Milky Way satellites. \hi-selected galaxy samples are inherently composed of fairly isolated systems as 21-cm surveys are not very efficient in probing high density regions \citep{Waugh2002a}. The galaxies that we probe in the tail of the velocity function are predominantly dwarf irregular galaxies and low mass spirals \citep[cf.][]{Zwaan2003a}.

To make a direct comparison with simulations, we need a \vc\ distribution function of dark matter haloes. \citet{Blanton2008a} compare an inferred velocity function of isolated dwarf galaxies with isolated haloes in the $N$-body, pure dark matter simulations of \citet{Kravtsov2004a}. 
Due to the first order similarity of the observed data sets, we also use this Kravtsov based sample as our dark matter halo reference distribution.

\begin{figure}
\begin{center}
\includegraphics[width=7.0cm,trim=1.2cm 4.3cm 1.5cm 0cm]{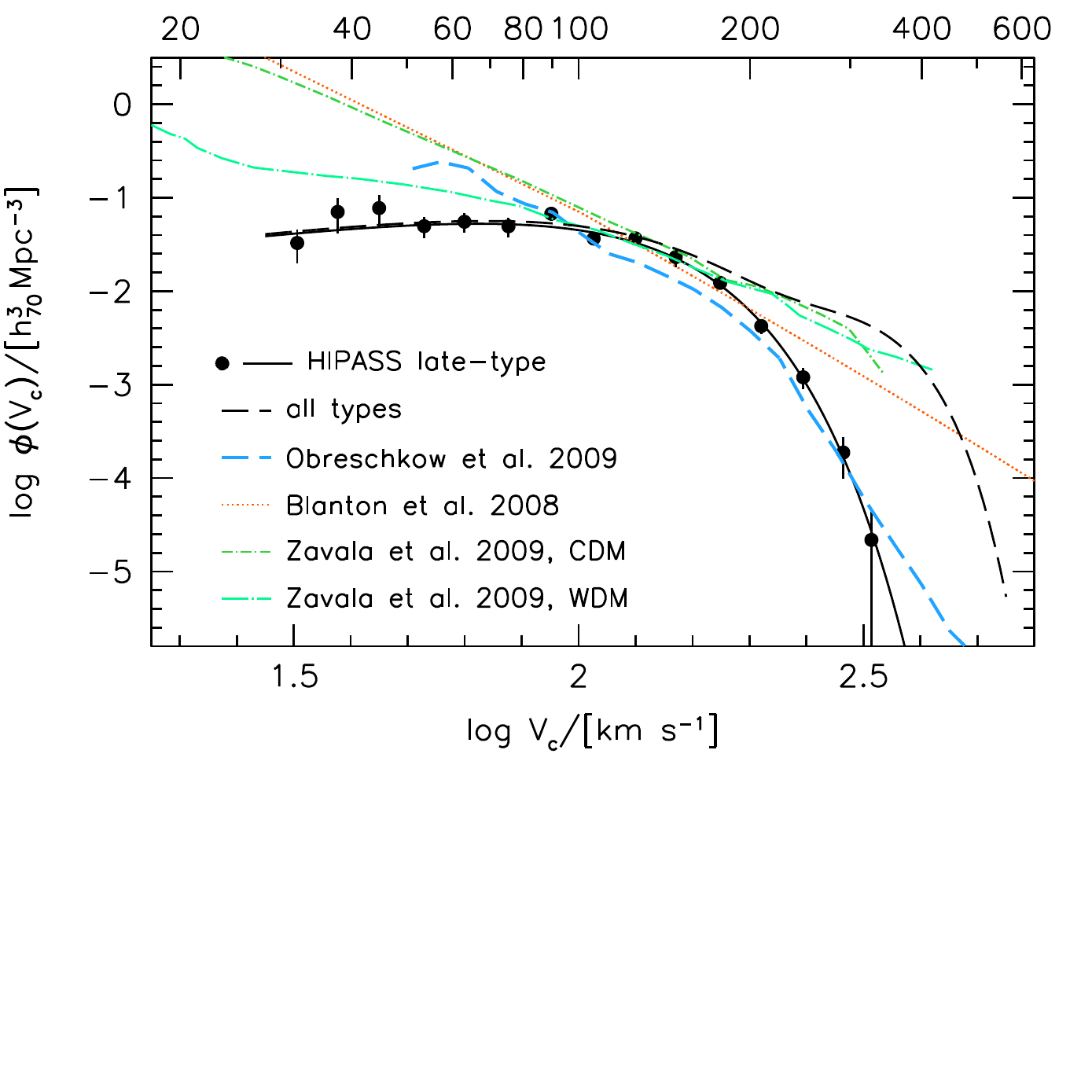}
\caption{\label{velf_theory.fig} The HIPASS velocity function (filled circles with solid line) compared with the velocity function from a selection of simulations. The blue dashed line shows the simulations of \citet{Obreschkow2009c} for late-type galaxies. The green  short dash-dotted and long dash-dotted lines are the CDM and WDM simulations from \citet{Zavala2009a} for all galaxy types. For comparison the observed velocity function for all galaxy types from Fig.\ref{velf_all.fig} is reproduced as the black dashed line. The red dotted line shows the dark matter haloes from \citet{Blanton2008a}. 
 }
\end{center}
\end{figure}

Unfortunately, \citet{Blanton2008a} present cumulative distribution functions of \vc, with makes quantitative comparisons very difficult. We differentiate their cumulative distribution function of dark matter haloes to derive $\phi(\vc)$ and show the result in Fig~\ref{velf_theory.fig} as a dotted line. Interestingly, the  space densities calculated from simulations and HIPASS match very well in the range $100 \,\kms < \vc <200\,\kms$. To some level, this is a coincidence because the definition of isolation for the dark matter haloes is not matched to that of the HIPASS catalogue.  \citet{Blanton2008a} show that the space densities in this \vc\ range increase somewhat if a more relaxed definition of isolation is applied.  Of more significance is the marked difference in the slope of the velocity function  below $\vc\approx 100\,\kms$, where the simulations predict a slope of $\alpha\approx -3$, compared to our observation of $\alpha\approx-1$. Simulations clearly predict many more haloes than are detected by HIPASS. 

There is one important caveat in this comparison, which is the mass resolution of the cosmological simulation. The \citep{Kravtsov2004a} simulations used by \citet{Blanton2008a} to calculate velocity functions, reliably determine space densities of haloes down to $\vc\approx 70\,\kms$. Below that value, we have used the power-law extrapolation of \citet{Blanton2008a}. 

Another interesting comparison with our observations is that with the recent results of \citet{Obreschkow2009c}. These authors have reprocessed the semi-analytic simulations of \citet{DeLucia2007a}, which, in turn, are based on the `millennium run', a cold dark matter (CDM) $N$-body simulation of a $(714 \,\rm Mpc/{h_{70}})^3$ volume, containing $3\times10^7$ galaxies. \citet{Obreschkow2009c} use a set of physical prescriptions to assign \hi\ and  
molecular hydrogen (H$_2$) to these galaxies and are able to reproduce many of the observed gas properties of local galaxies. The simulated \hi\ mass function is well matched to the one measured from HIPASS for \hi\ masses in excess of $\mhi\approx 2\times 10^8\msol$. The scatter in the observed and simulated $\mhi-W_{50}$ relation is very large (their Fig. 13), but this \mhi\ limit approximately translates into a rotational velocity limit of $\vc\approx 100\kms$, implying that space densities of galaxies below $\vc\approx 100\kms$ are not reliably predicted. 

To estimate \vc\ values from the \citet{Obreschkow2009c} catalogue, we use their measurements of $w_{HI}^{\rm peak}$, which is the line width between the two peaks of the double horned \hi\ profile. We use $\vc=(w_{HI}^{\rm peak}+12)/2$ to derive rotational velocities, where the correction of $12\,\kms$ is applied to correct for profile smoothing due to the velocity dispersion of the gas. That is, $w_{HI}^{\rm peak}$ slightly underestimates the true \vc\ as can be seen in Fig. 11 of \citet{Obreschkow2009c}.

The thick dashed line in Fig.~\ref{velf_theory.fig} shows the resulting velocity function, plotting the result for late-type galaxies to enable a direct comparison with the HIPASS results. The agreement in the shape of $\phi(\vc)$ above $\vc\approx 100\,\kms$ is striking, only above $\vc=300\,\kms$ do the simulations over-predict the observed number density. Unfortunately, at the low velocity end we cannot make a detailed comparison because the mass resolution of the millennium run is not sufficiently high to predict space densities in this regime. 

Higher resolution simulations are available from the recent work of \citet{Zavala2009a}. These authors use constrained cold and warm dark matter (WDM) simulations of a $(29 h_{70}^{-1} \,\rm Mpc)^3$ volume centered around the Local Group. The WDM simulations are based on a power spectrum using a fitting function that approximates the transfer function for a thermal WDM particle with $m_{\rm WDM}=1\rm keV$.
Their velocity functions are reproduced by short dash-dotted 
and long dash-dotted lines for CDM and WDM, respectively. Although their simulations assume that all galaxies are disk galaxies, they do not differentiate specifically between early or late type galaxies. We reproduce the velocity function for all galaxy types from Fig.\ref{velf_all.fig} as a long dashed line to compare the \citet{Zavala2009a} results with. The simulated curves are purported to be complete above $\vc=35\,\kms$. \citet{Zavala2009a} also test the effect of supernova feedback on the shape of the velocity function and conclude that it only marginally reduces the space densities at the low \vc\ end. 

For a comparison with their simulations, \citet{Zavala2009a} use preliminary results from the ALFALFA survey \citep{Giovanelli2005b}, concentrating on publicly released data bases in the Virgo direction region \citep{Giovanelli2007a} and the anti-Virgo direction region \citep{Saintonge2008a}. In total, the comparison is based on a small sample of 201 galaxies, 186 of which are in a high density region and the remaining 15 in a low density region. Only galaxies out to a distance of  $29 h_{70}^{-1} \,\rm Mpc$ are considered. Since the preliminary ALFALFA results are based on a much smaller galaxy sample, probing a smaller volume than HIPASS and concentrate only on a clear overdensity and an underdensity, we do not attempt to make a direct comparison with our results. We note, however, that the shape of the ALFALFA $\phi(\vc)$ as presented by \citet{Zavala2009a} is qualitatively in good agreement with our results.

Comparing our HIPASS measurements with the \citet{Zavala2009a} simulations, we essentially reach similar conclusions: above $\vc=100\,\kms$ the CDM predictions are in good agreement with the observations. Below  $\vc=100\,\kms$ CDM clearly predicts too many low mass haloes. Similar to the \citet{Blanton2008a} results, CDM predicts a steep low velocity slope as estimated by Eq.~\ref{slope.eq}. The WDM simulations predict a fairly flat velocity function, in much better agreement with the observations.

\section{Biases}

\begin{figure}
\begin{center}
\includegraphics[width=7.0cm,trim=1.2cm 4.3cm 1.5cm 0cm]{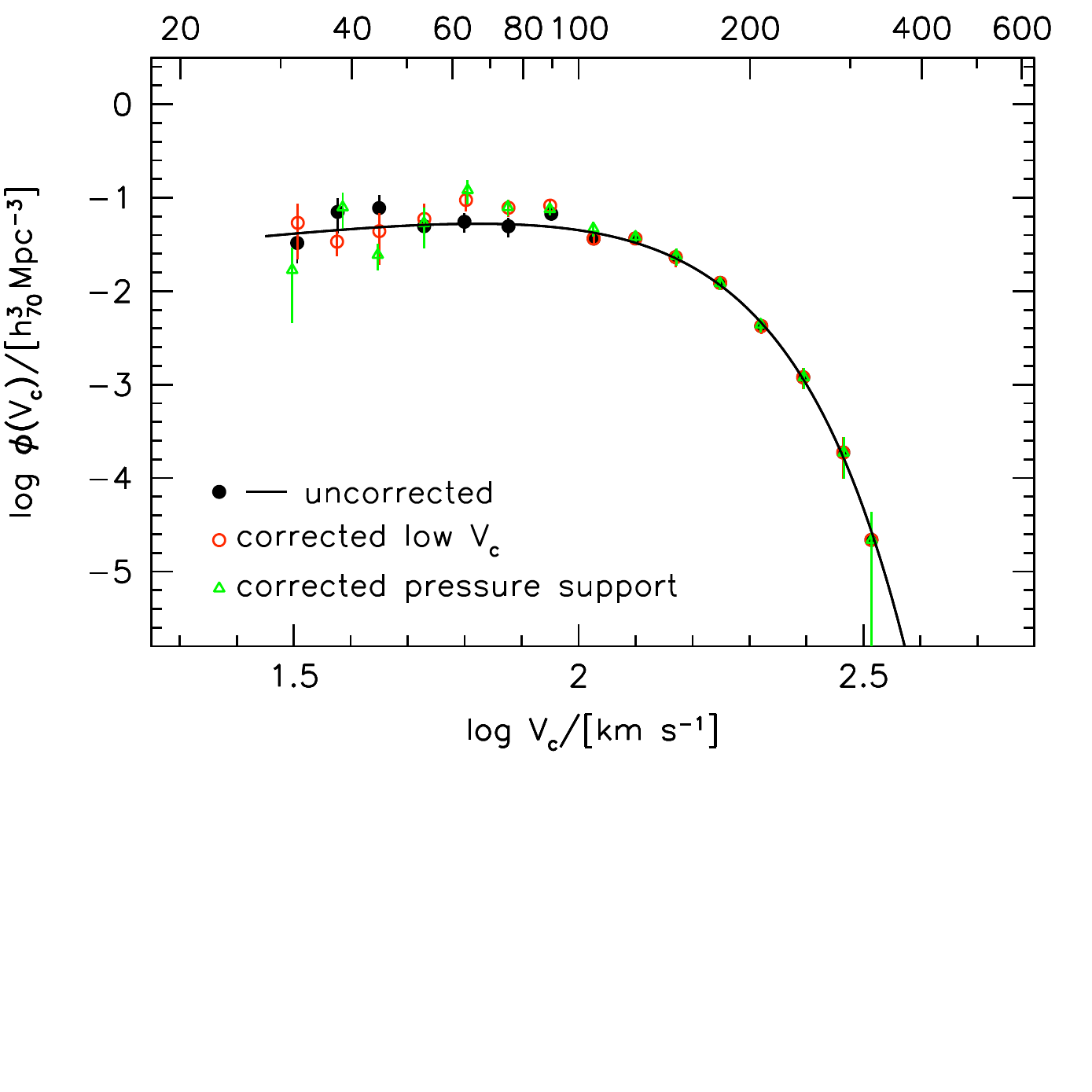}
\caption{\label{velf_bias2.fig} Testing the effect of possible biases on the shape of the rotational velocity function. The solid circles and line show the velocity function as presented in Fig.~\ref{velf_bias.fig}. The open circles demonstrate the effect of increasing the rotational velocity of low mass galaxies for which the flat part of the rotation curve might not be reached. The open triangles show how the velocity function changes if velocity widths are corrected for pressure support.}
\end{center}
\end{figure}

In making comparisons between our HIPASS results and simulations, there are a number of biases that should be discussed here. First, an important issue is whether the 21-cm velocity widths truly probe the full range of the galaxy velocity curve. If the detectable \hi\ is confined to the central regions of the galaxies, the \hi\ rotation curve might not measure the full rotational velocity of the dark matter halo \citep[e.g.,][]{Swaters2009a}. Inspecting the dwarf galaxy rotation curves of \citet{Swaters2009a}, it can be seen that most galaxies do just reach a flat part, but there are notable exceptions, especially those with rotational velocities lower than $\vc\approx 60\,\kms$. This implies that a fraction of our lower \vc\ measurements probably underestimates the true rotational velocity of the dark matter haloes. As an extreme test, we randomly multiply half our \vc\ measurements below $60 \,\kms$ with a factor 1.5 and construct $\phi(\vc)$ using this new estimates. Fig~\ref{velf_bias2.fig} shows the result, where filled symbols show the original space densities and open circles represent those after making the correction. Not surprisingly, the space density of objects with \vc\ just below 90 \kms\ is boosted after making the correction. In effect, the discrepancy between $\Lambda$CDM theory and our observations is slightly reduced just below the knee of the distribution, but not significantly at lower \vc. The effect that the space density drops below 50 \kms\ is a result of the fact that many \vc\ below that value shift to higher \vc\ bins, but no very low velocity systems replace them because they do not exist in our sample.

A second concern is whether gas velocity dispersions in low mass galaxies should be taken into account in calculating rotational velocities that measure the full potential. In very low mass galaxies, the gas velocity dispersions can be comparable in magnitude to the measured rotational velocities, adding a significant pressure support term compared to rotational support. If \hi\ synthesis maps are available, asymmetric drift correction can be made that correct for the pressure support. \citet{Begum2008b, Begum2008a} study a sample of very low mass galaxies with rotational velocities as low as 10 \kms. For each galaxy individually, they correct for pressure support and find significant corrections to \vc. We find that their corrections can be fitted as $\Delta\vc=130/\vc$, which we can apply to our data to make a rough correction for pressure support. The triangles in Fig~\ref{velf_theory.fig} show combined effect of this correction and the correction described in the previous paragraph.  We find that the pressure correction has no measurable effect on $\phi(\vc)$.

\begin{figure}
\begin{center}
\includegraphics[width=7.0cm,trim=1.2cm 4.3cm 1.5cm 0cm]{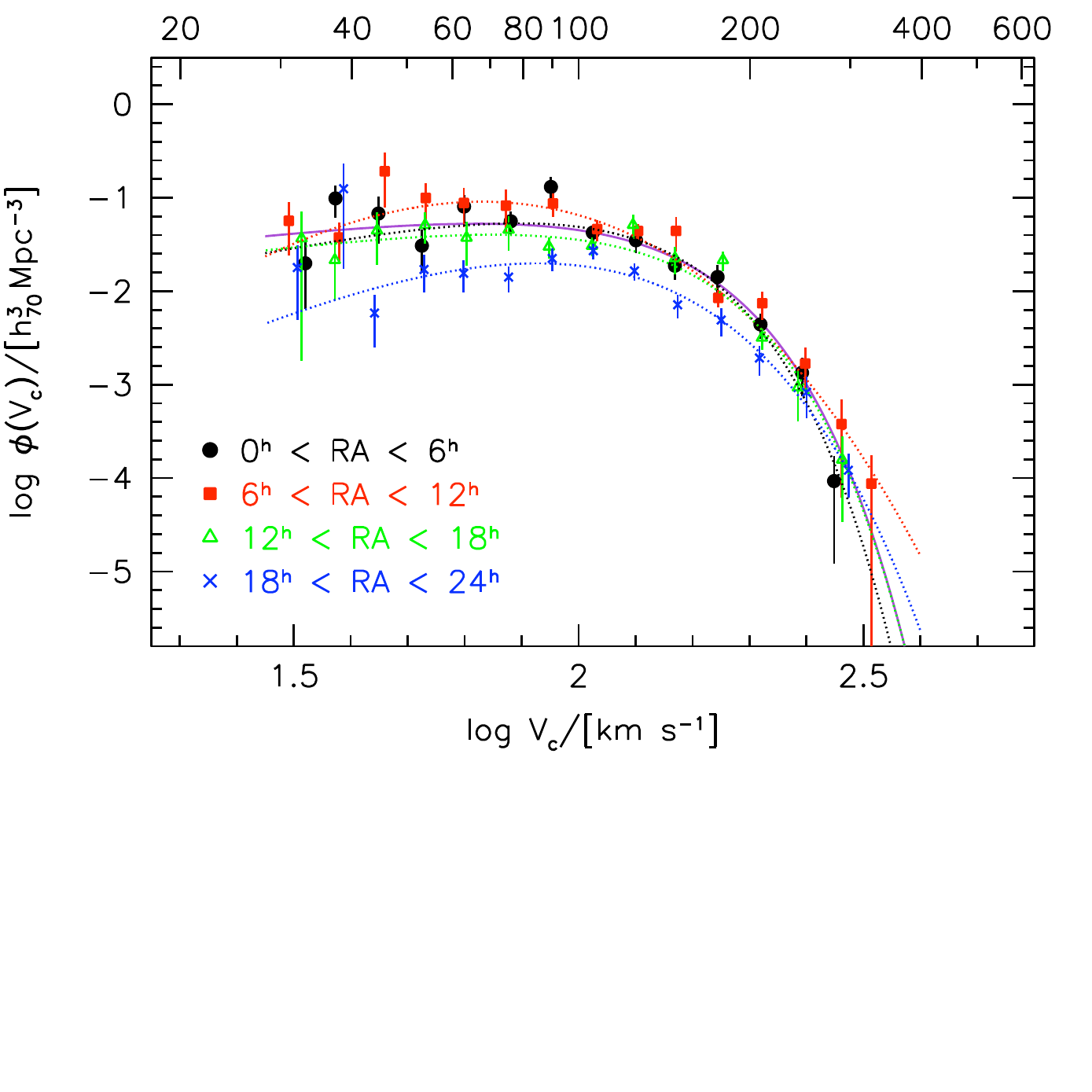}
\caption{\label{velf_cosvar.fig} The effect of cosmic variance on the circular velocity function. The four different symbols show the velocity function for four different quadrants of the southern sky, and dotted lines show modified Schechter fits. The solid line represents the fit to the full HIPASS sample.}
\end{center}
\end{figure}

Finally, we discuss how much cosmic variance can affect our measurements.   In Fig.~\ref{velf_cosvar.fig} we display the velocity function for four different quadrants of the southern sky. For each quadrant, we simply use the $\Sigma 1/{V_{\rm eff}}$ method as described in section 2, where the $V_{\rm eff}$ values are the maximum likelihood equivalents of the effective search volumes for each galaxy, calculated from the complete HIPASS sample. Also shown are the least-squares modified Schechter fits to each set of data points. Although the variation between the data points for the different quadrants is approximately a factor 2  at the high $\vc$ end and even larger at the low \vc\ tail, the fitted functions show good consistency between the quadrants. The only deviant velocity function is that of the fourth quadrant, which is clearly an underdense region. Despite a factor two lower overall normalization, the shape of the velocity function of the fourth quadrant is not significantly different from those of the other three quadrants. We conclude that our main result that the low \vc\ slope of the velocity function is 'flat' is unchanged if we subdivide our sample into smaller regions on the sky. Cosmic variance in the HIPASS survey volume is discussed in more detail in \citet{Zwaan2003a}.

\section{Summary and discussion}

We have used the HIPASS sample of \hi-selected galaxies to calculate the circular velocity function 
$\phi(\vc)$ of  late-type galaxies. Optical identifications of the HIPASS galaxies were used to determine
inclinations and morphological types. Our determination of the rotational velocity function is the first that
uses a direct method, where the same galaxy sample is used for measurements of \vc\ and the evaluation of space densities. Previous estimates of the  $\phi(\vc)$ were dependent on a conversion of the galaxy luminosity function and a relation between luminosity and rotational velocity (the Tully-Fisher relation). In these analyses different data sets were used for the calculation of the luminosity function and the Tully-Fisher relation. 

We find good agreement between our velocity function and inferred velocity function from the literature, but we are able to measure space densities down to $\vc=30\,\kms$, a factor of two lower than previous estimates. A standard Schechter function is not a good fit to $\phi(\vc)$  because space densities decline too rapidly at high velocities compared to an exponential drop-off. Instead, we use a modified Schechter function, which provides an excellent fit to the data points. In order to find a velocity function for all galaxy types, we add the early-type measurements from \citet{Chae2008a}.

As stated in the introduction, the velocity function is one of the most direct tests of galaxy formation models. Our velocity function is compared with those derived from $\Lambda$CDM galaxy formation simulations and we find a good agreement in the space density of objects with circular velocities $\vc>100\,\kms$. Below this value $\Lambda$CDM predicts too many objects, reaching a discrepancy of a factor 20 at $\vc=30\,\kms$.
WDM simulations are constructed to reduce the power on small scales and hence produce velocity functions with a flatter low-end slope, which provide a better fit to our data points. 
\citet{Primack2009a} argues that observational limits on the mass of the warm dark matter particle are already very strict. The circular velocity function essentially tests cosmological models on scales of approximately 1 Mpc \citep{Blanton2008a}. Similar scales are probed by the Ly$\alpha$ forest power spectrum and it is found that the mass of the WDM particle should be more than 4keV  \citep{Viel2008a} (or more than 1.7keV according to a more recent analysis by \citet{Boyarsky2009a}).  

\citet{Blanton2008a} use a sample of 12 low mass isolated dwarf galaxies and an independent measurement of the luminosity function to compare with $\Lambda$CDM predictions and claim consistency down to $\vc\approx 50 \kms$. Our analysis, and also that of \citet{Zavala2009a}, indicates that the inconsistency between observations and simulations is significant over a larger range in velocities, up to $\vc\approx 100\,\kms$. We argue that our constraints are firmer than those of \citet{Blanton2008a} because our HIPASS measurement is based on $\sim 50$ times more galaxies below $\vc=100\,\kms$  and is not reliant on optical luminosity functions. 

Apart from adopting WDM, which other factors could be responsible for the mismatch between observed and simulated rotational velocity functions? In section 5 we tested whether observational biases, such as dwarf galaxy rotation curves not probing the full potential or ignoring pressure terms could have an effect on $\phi(\vc)$ and concluded that the effect was insignificant. There are however astrophysical effects that could conceivably affect the shape of the measured $\phi(\vc)$. 

Our analysis is based on the assumption that low mass galaxies contain a detectable amount of neutral gas. Any gas-free low mass field galaxies would escape inclusion in the HIPASS sample. \citet{Geha2006a} obtained Green Bank Telescope 21cm observations of 101 low mass field galaxies and detected \hi\ emission in 88 per cent of those. It therefore seems unlikely that many low mass systems are missing from our analysis. A correction to the low \vc\ end of a factor $1/0.88$ would not change the slope significantly.

UV background photoionization of primordial gas could impact the formation of low mass objects \citep[e.g.,][]{Efstathiou1992a,Babul1992a}.  Recent calculations do not reach consistent conclusions on the magnitude of this effect and the characteristic \vc\ below which the effect becomes important. For example, variations in the reionization history, the inclusion of self-shielding in the calculations, and the link with SN feedback all have important effects \citep[e.g.,][and references therein]{Susa2004a,Hoeft2006a,Mesinger2008a,Sawala2009a,Pawlik2009a}. At present it remains unclear whether the suppression of low mass system formation by the UV background is sufficient to explain the observed local velocity function. Similarly, the effect of SN feedback might be important. SN feedback implies chemical enrichment, heating and entraining of surrounding gas, reducing the gas fractions of low mass systems \citep[e.g.][]{Dekel1986a,White1991a}. The treatment of SN feedback in simulations of galaxy formation is very complicated and, like UV photoionization, the importance of the effect is still a topic of discussion \citep{Scannapieco2006a}. It should be noted though that \citet{Zavala2009a} conclude that the SN feedback has only a minor effect on the $z=0$ circular velocity function.

Future blind \hi\ surveys with instruments such as the Australian SKA Pathfinder (ASKAP) will probe much larger volumes than HIPASS with a lower noise level and much higher spatial resolution. For instance, a one year survey covering 75 per cent of the sky will find approximately $5\times 10^5$ galaxies, $3\times 10^4$ of which will be spatially resolved, thus allowing the measurement of rotation curves and hence of more accurate rotational velocities. This, in combination with the much larger number of galaxies, a higher sensitivity to low mass galaxies, and the ability to study different environments, will allow much more accurate  measurements of the galaxy rotational velocity function in the local universe.

\section*{Acknowledgments}
MAZ acknowledges the hospitality of  the University of Western Australia in Perth where most of this work was carried out. We thank the anonymous referee for a detailed and constructive report.
The Parkes telescope is part of the Australia Telescope which is funded by the Commonwealth of Australia for operation as a National Facility managed by CSIRO.

\small
\bibliographystyle{mnras}
\bibliography{/Users/mzwaan/REFERENCES/zwaanreferences}

\label{lastpage}

\end{document}